\documentclass[aps,prl,twocolumn,showpacs,superscriptaddress,nofootinbib]{revtex4}

\usepackage{amsmath}   
\usepackage{xspace}
\usepackage{units}
\usepackage{graphicx}  
\usepackage[caption=false]{subfig}

\usepackage{hyperref}

\newcommand{\minerva}{MINERvA\xspace}

\newcommand{\genie}{\textsc{genie}\xspace}

\newcommand{\tcoh}{\ensuremath{\left| t\right| \xspace}}

\newcommand{\numu}{\ensuremath{\nu_{\mu}}\xspace}

\newcommand{\pip}{\ensuremath{\pi^{+}}\xspace}

\newcommand{\piz}{\ensuremath{\pi^{0}}\xspace}
\newcommand{\lam}{\ensuremath{\Lambda}\xspace}
\newcommand{\sig}{\ensuremath{\Sigma}\xspace}

\newcommand{\kp}{\ensuremath{K^{+}}\xspace}

\newcommand{\kz}{\ensuremath{K^{0}}\xspace}

\newcommand{\kzS}{\ensuremath{K^{0}_{\mathrm{S}}}\xspace}

\newcommand{\mum}{\ensuremath{\mu^{-}}\xspace}

\newcommand{\sizecheck}{0} 
\newcommand{\PRLsupp}{0}   
\ifnum\PRLsupp=0

\else

\fi

\newif\ifpdf
\ifx\pdfoutput\undefined
   \pdffalse
\else
   \pdfoutput=1
   \pdftrue
\fi
\ifpdf
   \usepackage{graphicx}
   \usepackage{epstopdf}
\else
   \usepackage{graphicx}
\fi

\begin{document}

\title{First evidence of coherent $K^{+}$ meson production in neutrino-nucleus scattering}



\newcommand{\deceased}{Deceased}

\newcommand{\Rutgers}{Rutgers, The State University of New Jersey, Piscataway, New Jersey 08854, USA}
\newcommand{\Hampton}{Hampton University, Dept. of Physics, Hampton, VA 23668, USA}
\newcommand{\Dortmund}{Institute of Physics, Dortmund University, 44221, Germany }
\newcommand{\Otterbein}{Department of Physics, Otterbein University, 1 South Grove Street, Westerville, OH, 43081 USA}
\newcommand{\JMU}{James Madison University, Harrisonburg, Virginia 22807, USA}
\newcommand{\Florida}{University of Florida, Department of Physics, Gainesville, FL 32611}
\newcommand{\UCIrvine}{Department of Physics and Astronomy, University of California, Irvine, Irvine, California 92697-4575, USA}
\newcommand{\CBPF}{Centro Brasileiro de Pesquisas F\'{i}sicas, Rua Dr. Xavier Sigaud 150, Urca, Rio de Janeiro, Rio de Janeiro, 22290-180, Brazil}
\newcommand{\PUCP}{Secci\'{o}n F\'{i}sica, Departamento de Ciencias, Pontificia Universidad Cat\'{o}lica del Per\'{u}, Apartado 1761, Lima, Per\'{u}}
\newcommand{\INRM}{Institute for Nuclear Research of the Russian Academy of Sciences, 117312 Moscow, Russia}
\newcommand{\Jlab}{Jefferson Lab, 12000 Jefferson Avenue, Newport News, VA 23606, USA}
\newcommand{\Pittsburgh}{Department of Physics and Astronomy, University of Pittsburgh, Pittsburgh, Pennsylvania 15260, USA}
\newcommand{\Guanajuato}{Campus Le\'{o}n y Campus Guanajuato, Universidad de Guanajuato, Lascurain de Retana No. 5, Colonia Centro, Guanajuato 36000, Guanajuato M\'{e}xico.}
\newcommand{\Athens}{Department of Physics, University of Athens, GR-15771 Athens, Greece}
\newcommand{\Tufts}{Physics Department, Tufts University, Medford, Massachusetts 02155, USA}
\newcommand{\WM}{Department of Physics, College of William \& Mary, Williamsburg, Virginia 23187, USA}
\newcommand{\FNAL}{Fermi National Accelerator Laboratory, Batavia, Illinois 60510, USA}
\newcommand{\Purdue}{Department of Chemistry and Physics, Purdue University Calumet, Hammond, Indiana 46323, USA}
\newcommand{\MCLA}{Massachusetts College of Liberal Arts, 375 Church Street, North Adams, MA 01247}
\newcommand{\UMD}{Department of Physics, University of Minnesota -- Duluth, Duluth, Minnesota 55812, USA}
\newcommand{\Northwestern}{Northwestern University, Evanston, Illinois 60208}
\newcommand{\UNI}{Universidad Nacional de Ingenier\'{i}a, Apartado 31139, Lima, Per\'{u}}
\newcommand{\Rochester}{University of Rochester, Rochester, New York 14627 USA}
\newcommand{\Austin}{Department of Physics, University of Texas, 1 University Station, Austin, Texas 78712, USA}
\newcommand{\USM}{Departamento de F\'{i}sica, Universidad T\'{e}cnica Federico Santa Mar\'{i}a, Avenida Espa\~{n}a 1680 Casilla 110-V, Valpara\'{i}so, Chile}
\newcommand{\Geneva}{University of Geneva, 1211 Geneva 4, Switzerland}
\newcommand{\Chicago}{Enrico Fermi Institute, University of Chicago, Chicago, IL 60637 USA}
\newcommand{\hired}{}
\newcommand{\OregonState}{Department of Physics, Oregon State University, Corvallis, Oregon 97331, USA}
\newcommand{\oxford}{}
\newcommand{\bmeThanks}{now at SLAC National Accelerator Laboratory, Stanford, CA 94309, USA}
\newcommand{\higueraThanks}{now at University of Houston, Houston, TX 77204, USA}
\newcommand{\damartinezThanks}{now at Illinois Institute of Technology, Chicago, IL 60616, USA}
\newcommand{\mcgivernThanks}{now at Iowa State University, Ames, IA 50011, USA}
\newcommand{\joelmousseauThanks}{now at University of Michigan, Ann Arbor, MI 48109, USA}
\newcommand{\LazaThanks}{also at Department of Physics, University of Antananarivo, Madagascar}
\newcommand{\twaltonThanks}{now at Fermi National Accelerator Laboratory, Batavia, IL 60510, USA}
\newcommand{\jwolcottThanks}{now at Tufts University, Medford, MA 02155, USA}
\newcommand{\gzgThanks}{Deceased}

\author{Z.~Wang}                          \affiliation{\Rochester}
\author{C.M.~Marshall}                    \affiliation{\Rochester}

\author{L.~Aliaga}                        \affiliation{\WM}  \affiliation{\PUCP}
\author{O.~Altinok}                       \affiliation{\Tufts}
\author{L.~Bellantoni}                    \affiliation{\FNAL}
\author{A.~Bercellie}                     \affiliation{\Rochester}
\author{M.~Betancourt}                    \affiliation{\FNAL}
\author{A.~Bodek}                         \affiliation{\Rochester}
\author{A.~Bravar}                        \affiliation{\Geneva}
\author{H.~Budd}                          \affiliation{\Rochester}
\author{T.~Cai}                           \affiliation{\Rochester}
\author{M.F.~Carneiro}                    \affiliation{\CBPF}
\author{H.~da~Motta}                      \affiliation{\CBPF}
\author{S.A.~Dytman}                      \affiliation{\Pittsburgh}
\author{G.A.~D\'{i}az~}                   \affiliation{\Rochester}  \affiliation{\PUCP}
\author{B.~Eberly}\thanks{\bmeThanks}     \affiliation{\Pittsburgh}
\author{E.~Endress}                       \affiliation{\PUCP}
\author{J.~Felix}                         \affiliation{\Guanajuato}
\author{L.~Fields}                        \affiliation{\FNAL}  \affiliation{\Northwestern}
\author{R.~Fine}                          \affiliation{\Rochester}
\author{R.Galindo}                        \affiliation{\USM}
\author{H.~Gallagher}                     \affiliation{\Tufts}
\author{A.~Ghosh}                         \affiliation{\USM}  \affiliation{\CBPF}
\author{T.~Golan}                         \affiliation{\Rochester}  \affiliation{\FNAL}
\author{R.~Gran}                          \affiliation{\UMD}
\author{D.A.~Harris}                      \affiliation{\FNAL}
\author{A.~Higuera}\thanks{\higueraThanks}  \affiliation{\Rochester}  \affiliation{\Guanajuato}
\author{K.~Hurtado}                       \affiliation{\CBPF}  \affiliation{\UNI}
\author{M.~Kiveni}                        \affiliation{\FNAL}
\author{J.~Kleykamp}                      \affiliation{\Rochester}
\author{M.~Kordosky}                      \affiliation{\WM}
\author{T.~Le}                            \affiliation{\Tufts}  \affiliation{\Rutgers}
\author{E.~Maher}                         \affiliation{\MCLA}
\author{S.~Manly}                         \affiliation{\Rochester}
\author{W.A.~Mann}                        \affiliation{\Tufts}
\author{D.A.~Martinez~Caicedo}\thanks{\damartinezThanks}  \affiliation{\CBPF}
\author{K.S.~McFarland}                   \affiliation{\Rochester}  \affiliation{\FNAL}
\author{C.L.~McGivern}\thanks{\mcgivernThanks}  \affiliation{\Pittsburgh}
\author{A.M.~McGowan}                     \affiliation{\Rochester}
\author{B.~Messerly}                      \affiliation{\Pittsburgh}
\author{J.~Miller}                        \affiliation{\USM}
\author{A.~Mislivec}                      \affiliation{\Rochester}
\author{J.G.~Morf\'{i}n}                  \affiliation{\FNAL}
\author{J.~Mousseau}\thanks{\joelmousseauThanks}  \affiliation{\Florida}
\author{D.~Naples}                        \affiliation{\Pittsburgh}
\author{J.K.~Nelson}                      \affiliation{\WM}
\author{A.~Norrick}                       \affiliation{\WM}
\author{Nuruzzaman}                       \affiliation{\Rutgers}  \affiliation{\USM}
\author{V.~Paolone}                       \affiliation{\Pittsburgh}
\author{J.~Park}                          \affiliation{\Rochester}
\author{C.E.~Patrick}                     \affiliation{\Northwestern}
\author{G.N.~Perdue}                      \affiliation{\FNAL}  \affiliation{\Rochester}
\author{L.~Rakotondravohitra}\thanks{\LazaThanks}  \affiliation{\FNAL}
\author{M.A.~Ramirez}                     \affiliation{\Guanajuato}
\author{R.D.~Ransome}                     \affiliation{\Rutgers}
\author{H.~Ray}                           \affiliation{\Florida}
\author{L.~Ren}                           \affiliation{\Pittsburgh}
\author{D.~Rimal}                         \affiliation{\Florida}
\author{P.A.~Rodrigues}                   \affiliation{\Rochester}
\author{D.~Ruterbories}                   \affiliation{\Rochester}
\author{H.~Schellman}                     \affiliation{\OregonState}  \affiliation{\Northwestern}
\author{D.W.~Schmitz}                     \affiliation{\Chicago}  \affiliation{\FNAL}
\author{C.~Simon}                         \affiliation{\UCIrvine}
\author{C.J.~Solano~Salinas}              \affiliation{\UNI}
\author{B.G.~Tice}                        \affiliation{\Rutgers}
\author{E.~Valencia}                      \affiliation{\Guanajuato}
\author{T.~Walton}\thanks{\twaltonThanks}  \affiliation{\Hampton}
\author{J.~Wolcott}\thanks{\jwolcottThanks}  \affiliation{\Rochester}
\author{M.Wospakrik}                      \affiliation{\Florida}
\author{G.~Zavala}\thanks{\gzgThanks}     \affiliation{\Guanajuato}
\author{D.~Zhang}                         \affiliation{\WM}

\collaboration{\minerva  Collaboration}\ \noaffiliation

\date{\today}

\pacs{13.15.+g,25.30.Pt}
\begin{abstract}
Neutrino-induced charged-current coherent kaon production,
$\nu_{\mu}A\rightarrow\mu^{-}K^{+}A$, is a rare, inelastic electroweak 
process that brings a $K^+$ on shell and leaves the
target nucleus intact in its ground state.  This process is significantly
lower in rate than neutrino-induced charged-current coherent pion
production, because of Cabibbo suppression and a kinematic suppression
due to the larger kaon mass.  We search for such events in the
scintillator tracker of MINERvA by observing the final state $K^+$,
$\mu^-$ and no other detector activity, and by using the kinematics of the
final state particles to reconstruct the small momentum transfer to
the nucleus, which is a model-independent characteristic of coherent scattering. 
We find the first experimental evidence for the process at $3\sigma$ significance.
\end{abstract}
\ifnum\sizecheck=0  
\maketitle
\fi

\begin{figure}[bt]	
\centering
\ifnum\PRLsupp=0
  \includegraphics[width=0.9\columnwidth]{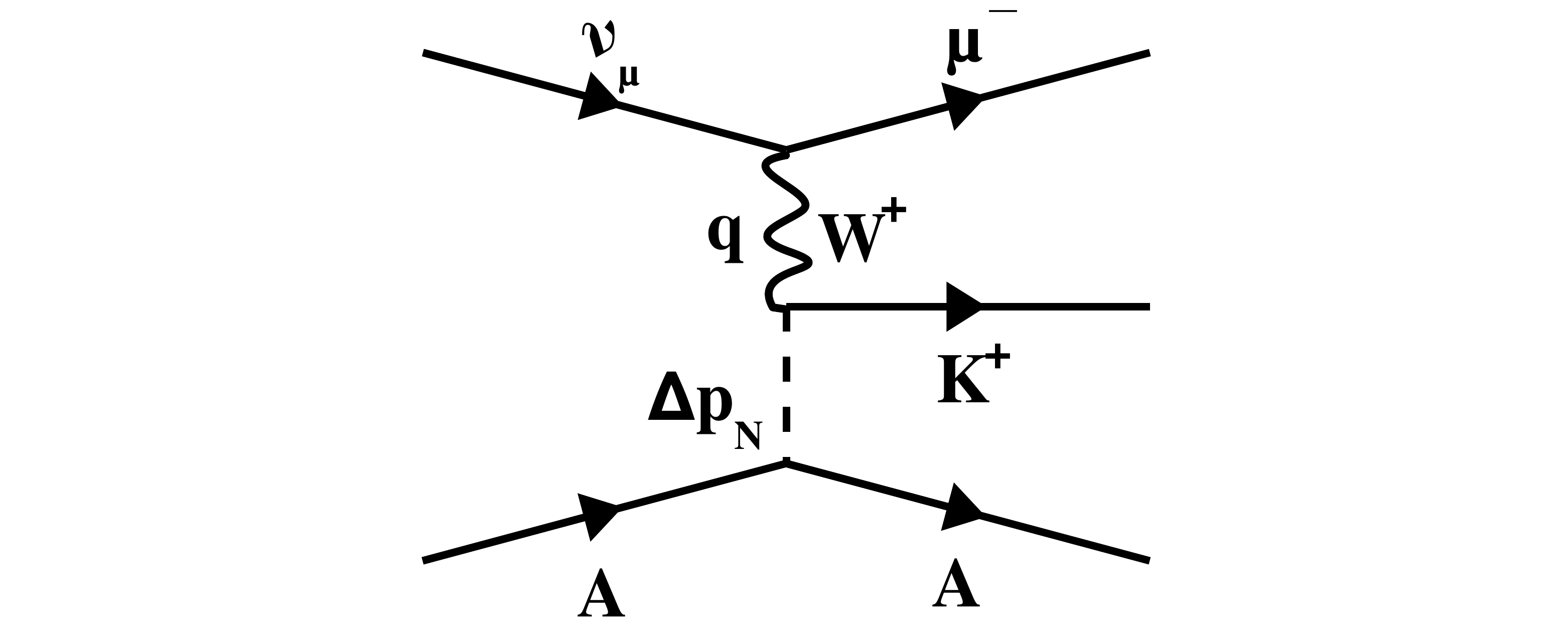}
\else
  \includegraphics[width=0.9\columnwidth]{CohDiagram_wide.pdf}
\fi

\caption{Feynman diagram for coherent charged kaon production. The square of the momentum transfer to the nucleus is $|\Delta p_N|^2 = |q-p_K|^2 = \tcoh$.}
\label{fig:feyn-coh}
\end{figure}

Charged mesons may be produced in inelastic, coherent
charged-current reactions of neutrinos. This 
reaction is believed to occur when an off-shell $W$ boson fluctuates
to a meson. The meson is brought on the mass shell by exchange of a
particle carrying no quantum numbers with the target nucleus, as
illustrated in Fig.~\ref{fig:feyn-coh}.  Charged pion production
through this mechanism has been observed and measured with
$\mathcal{O}$(10)\% precision on nuclei ranging from carbon to argon
and across a range of neutrino (and antineutrino) energies from a few GeV to tens of
GeV~\cite{Grabosch:1985mt,Marage:1986cy,Allport:1988cq,Vilain:1993sf,Hasegawa:2005td,Hiraide:2008eu,PhysRevLett.113.261802,Acciarri:2014eit}.
If the mechanism described above is the source of these events, then
there should be an analogous, Cabibbo-suppressed process in which a
$K^\pm$ meson is produced. In this Letter, we report the first experimental
evidence of this process.

The exchange with the nucleus must leave the nucleus in its ground
state for the process to be coherent.  This requires that the four
momentum transfer to the nucleus, $\Delta p_N\equiv p_{Af}-p_{Ai}$,
satisfy the relation $\left| t\right|\equiv \left| (\Delta
p_N)^2 \right| \leq\hbar^{2}/R^{2}$, where $R$ is the radius of the
nucleus.
Adler's theorem~\cite{Adler:1964yx} relates the coherent meson
production cross section at $Q^2\equiv-q^2=0$ to the meson-nucleus
elastic cross section~\cite{Piketty:1970sq,Lackner:1979ax,Rein:1982pf}.
In the limit of muon and meson masses $m_\mu,m_M\ll E_\nu$, 
where $E_{\nu}$ is the neutrino energy,

\begin{equation}\label{eqn:coh-0Q2}
\frac{d^3\sigma_{\mathrm coh}}{dQ^2dy\,d\!\tcoh}\Bigg{|}_{Q^2=0}
=\frac{G_F^2}{2\pi^2}\,f_M^2\,\frac{1-y}{y}\,\frac{d\sigma(M A\to M A)}{d\!\tcoh},
\end{equation}

\noindent
where $E_M$ is the final state meson energy, $y=E_M/E_\nu$, and $f_M$
is the meson decay constant.  The meson-nucleus elastic cross section
and its exponential decrease with increasing \tcoh\ are parameterized
from meson-nucleus scattering
data~\cite{Lackner:1979ax,Rein:1982pf,Berger:2008xs,Andreopoulos201087,
Hayato:2009zz,Casper:2002sd,Golan:2012wx}.  Models must be used to
extrapolate away from $Q^2 = 0$.  The model most commonly used in neutrino
event generators~\cite{Andreopoulos201087, Hayato:2009zz,
Casper:2002sd,Golan:2012wx} is that of Rein and
Sehgal~\cite{Rein:1982pf}, which assumes no vector current and
extrapolates the axial-vector current using a multiplicative dipole
form factor $F^2_{dipole}(Q^2)=1/(1+Q^2/m_A^2)^2$ to modify
Eq.~\ref{eqn:coh-0Q2}.  Other authors have proposed alternate
extrapolations away from $Q^2 = 0$~\cite{Gershtein:1980vd,Belkov:1986hn,Berger:2008xs,Paschos:2009ag}.
At low energies, modifications to Eq.~\ref{eqn:coh-0Q2} due to finite
masses become important, restricting the allowed kinematics to $Q^2\ge m^2_\mu\frac{y}{1-y}$
and $\tcoh\ge\left( \frac{Q^2+m_M^2}{2yE_\nu}\right)
^2$~\cite{Rein:2006di,Higuera:2013pra}. This restriction on \tcoh~removes more 
phase space for kaon production than it does for pion
production due to the larger meson mass, $m_M$.  An alternate approach
is to start with the Cabibbo-suppressed single kaon production cross section on nucleons~\cite{PhysRevD.82.033001} 
at low neutrino energies and calculate a coherent sum~\cite{AlvarezRuso:2012fc}. 

Standard neutrino interaction generators~\cite{Andreopoulos201087,
Hayato:2009zz, Casper:2002sd,Golan:2012wx} have
no model for coherent \kp production.  Even in PCAC models,
the dearth of data on low energy $K^+$-nucleus elastic scattering
at the relevant kaon energies~\cite{Gobbi:1972rq,
Sternheim:1974ct} precludes a precise
calculation.  However, the rate in the signal model does not affect
the result, and in fact the model is only needed to determine
energy deposited near the interaction point and the distribution of \tcoh,
and both of those quantities are most affected by well-modeled detector
resolutions. Therefore, we generate coherent pion production with \genie
2.8.4~\cite{Andreopoulos201087} and recalculate hadronic kinematics while holding the lepton kinematics
and magnitude of the four-momentum transfer to the nucleus fixed.  The larger
kaon mass means there is only a physical solution for 26\% of pion events for
MINERvA's neutrino flux.  The rate is also reduced by a factor of
$\left(f_K/f_\pi \right)^2 \tan^2\theta_C = 0.077\pm0.001$~\cite{PDG}, where $\theta_C$
is the Cabibbo angle. The ratio of the \kp to \pip elastic scattering cross sections
on carbon is calculated to be $\sim0.7$~\cite{boris}, consistent with measurements at
higher meson energies~\cite{Gobbi:1972rq}. We therefore expect our flux-
averaged cross section for coherent \kp production to be $\sim 4 \times 10^{−41}$~cm$^2$ per
carbon nucleus.

The signal reaction produces events with a forward $K^+$ and $\mu^-$
and no additional energy near the vertex.  Processes that produce the
same final state, but often with additional low energy particles at
the interaction vertex and with large \tcoh, are simulated using the
\genie 2.8.4 neutrino event generator~\cite{Andreopoulos201087}.
Inelastic reactions for
$W < 1.7$~GeV are simulated with a tuned model of discrete baryon
resonance production~\cite{Rein:1980wg}. The transition to deep
inelastic scattering is simulated using the Bodek-Yang
model~\cite{Bodek:2004pc}. Hadronization at higher energies is
simulated with the AGKY model~\cite{Yang:2009zx} based on the gradual
transition from KNO scaling~\cite{Koba:1972ng} to the LUND string model
of PYTHIA~\cite{Sjostrand:2006za} with increasing final state hadronic invariant mass, $W$. 
In \genie, parameters that control the rate of
strange particle production in hadronization are tuned such that rates
of \lam and \kzS production on deuterium and neon agree with
bubble chamber measurements~\cite{Jones:1992bm, Allasia:1983vd, Bosetti:1982vk,
Willocq:1991dy,Baker:1986xx, DeProspo:1994ac}. Final state interactions, in which
hadrons interact within the target nucleus, are modeled using the
INTRANUKE package~\cite{Andreopoulos201087}. \genie does not simulate FSI  
processes where \kp are produced in \pip interactions, for example 
$\pip n \rightarrow \kp \lam$. However, \pip strong interactions can 
only produce pairs of strage particles, and \lam or \sig baryons will 
cause events to be rejected.


Cabibbo-suppressed $\Delta S = 1$ reactions are an important
background not modeled in \genie 2.8.4. In particular, single \kp\ 
production off a bound neutron, $\numu n \rightarrow \mum \kp n$, has
the same apparent signature as the coherent reaction if the neutron does not
interact near the vertex and if the reaction happens to reconstruct to
low \tcoh.  
To include the $\Delta S = 1$ background,
we simulate the single kaon process using \genie
2.10.0~\cite{Alam:2015x} based on the model described in Ref.~\cite{PhysRevD.82.033001}. 
To preserve the total \kp production cross
section, $\Delta S = 0$ events are removed using a parameterization of 
the $\Delta S = 1$ to $\Delta S = 0$ cross section ratio as a function of $W$~\footnote{An emperical fit gives $\frac{\sigma_{\Delta S = 1}(W)}{\sigma_{\Delta S = 0}(W)} \approx -0.016 + \frac{0.28}{W^{1.09}}$}.

This measurement uses data taken by \minerva in the NuMI 
beamline~\cite{Adamson:2015dkw} at Fermilab. 
The data used in
this analysis were taken between March 2010 and April 2012 and correspond to 
$3.33 \times 10^{20}$ protons on target, in a $\nu_\mu$-enriched beam with a peak neutrino energy of 3.5~GeV.
A Geant4-based model is used to simulate the neutrino
beam. This model is constrained to reproduce thin-target hadron production
measurements on carbon by the NA49~\cite{Alt:2006fr} and MIPP~\cite{Lebedev:2007zz}
experiments. The 8.5\% uncertainty on
the prediction of the neutrino flux is set by the precision in these
measurements, and the uncertainties in the beamline
focusing system~\cite{Pavlovic:2008zz}. 

The \minerva detector is described in Ref.~\cite{minerva_nim}. For this result, the
interaction vertex is constrained to be within a 5.57 metric ton volume of plastic scintillator, consisting
of 95\% CH and 5\% other materials by mass. The MINOS near detector is
a magnetized iron spectrometer~\cite{Michael:2008bc} located 2~m
downstream of \minerva and is used to reconstruct the
momentum and charge of $\mu^{\pm}$.

We consider events with exactly two charged particle tracks originating from the neutrino
interaction point: one \mum and one \kp candidate. The \mum candidate must
exit the back of \minerva and match to a negatively-charged track entering
the front of MINOS. Timing information is used to identify delayed decay products near the
endpoint of the \kp candidate, consistent with the 12.4~ns \kp lifetime.
The \kp identification algorithm is described in detail
in Ref.~\cite{Marshall:2016}.

\begin{figure}[bt]	
\centering
\mbox{   \subfloat[]{\includegraphics[width=0.5\columnwidth,bb=20 20 575 520]{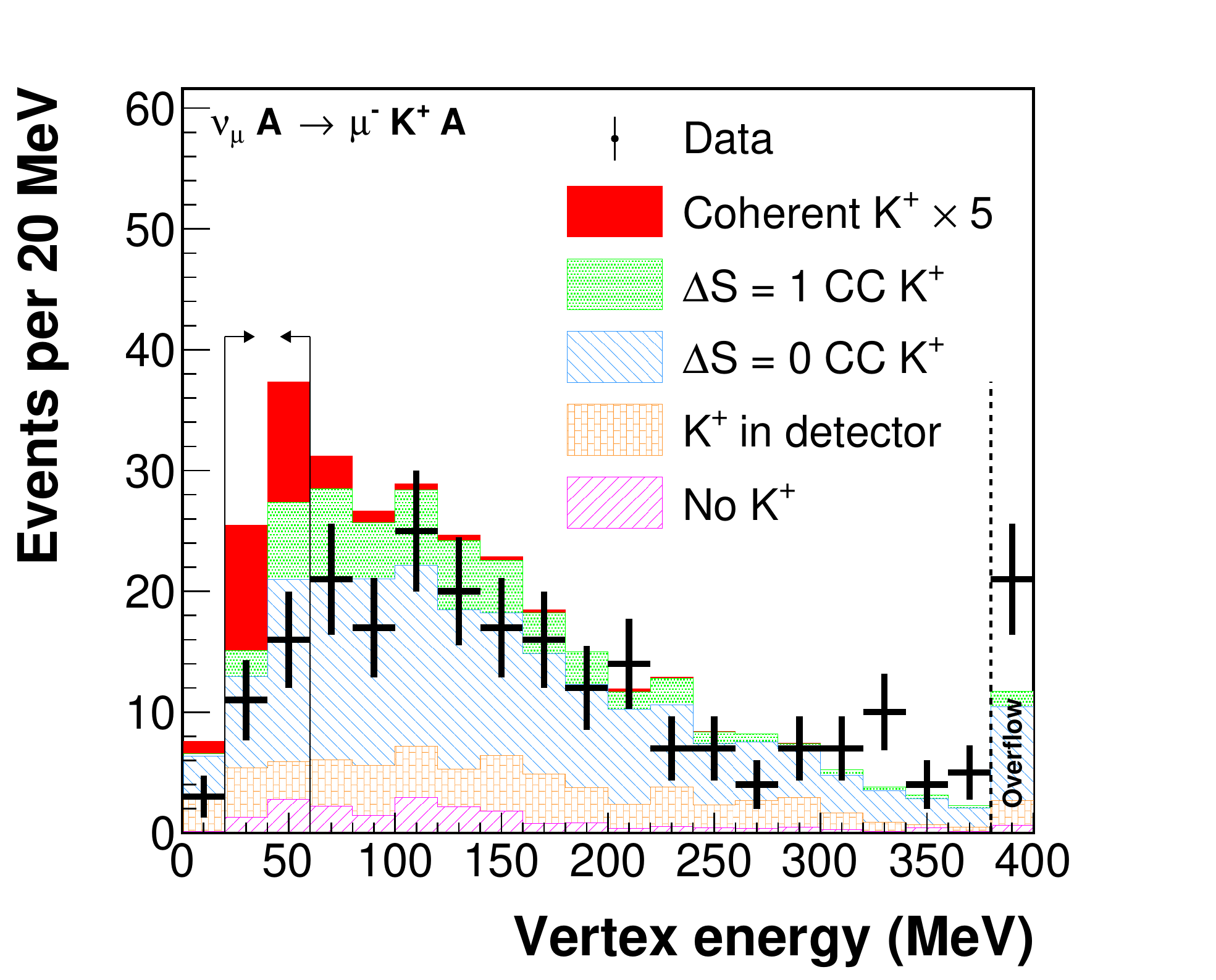} \label{fig:vtx}}
         \subfloat[]{\includegraphics[width=0.5\columnwidth,bb=20 20 575 554]{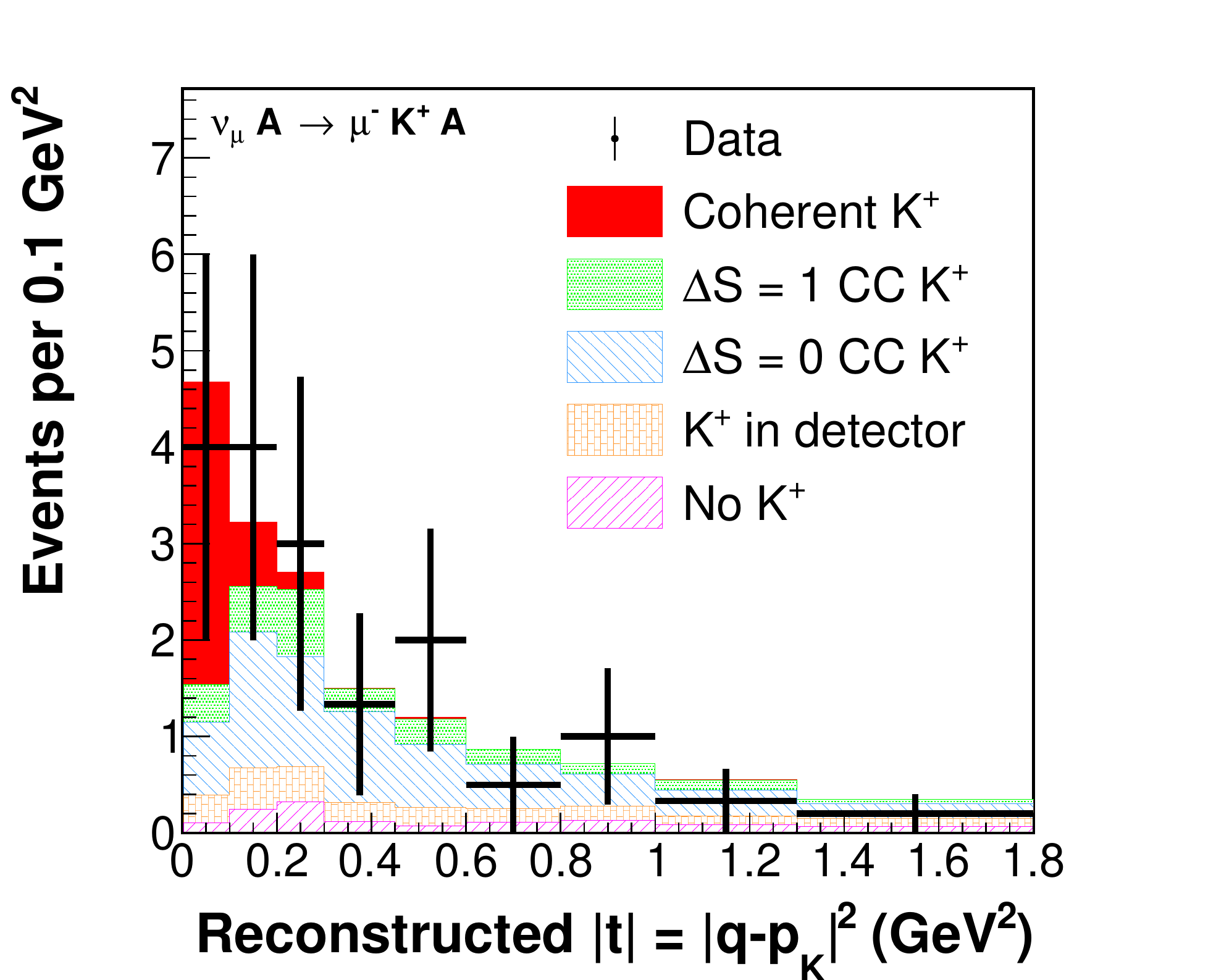} \label{fig:t}}}
\caption{(a) The distribution of vertex energy before sideband tuning. The signal simulation is enhanced by a factor of 5. (b) The distribution of \tcoh\ after selection on vertex energy and tuning of the background components, but before the removal of events with prompt $\pi^0$s. In both figures, data points have statistical uncertainties only.}
\end{figure}

Events with evidence of nuclear break-up are rejected by
considering energy deposited in a region around the interaction
point. This ``vertex region'' extends 10~cm in each direction along the
detector axis (3.5$^\circ$ above the beam direction), and 20~cm in the transverse direction. In signal
reactions, energy in this region is due to the muon and kaon
only. Additional charged hadrons
deposit more energy near the vertex. Events are
selected when the energy in the vertex region, $E_{vtx}$, is between 20 and 60
MeV, as shown in Fig.~\ref{fig:vtx}. For the two-track sample, 75\% of coherent events are
retained by the vertex energy cut and 85\% of two-track backgrounds are rejected.
The prediction from simulation exceeds the data at very low vertex energy, 
consistent with previous \minerva
results~\cite{PhysRevLett.113.261802,nuprl,PhysRevLett.116.071802}, which require
additional events with multiple nucleons beyond the \genie prediction.

The kaon energy is measured calorimetrically by summing all energy in \minerva not
associated with the muon track. For coherent events, there are no other final state
particles and this energy is due to the \kp and the 
products of its interactions in the detector. A calorimetric factor of 1.2 is
applied so that the kaon energy residual in the signal simulation is
peaked at zero. The reconstructed neutrino energy, $E_{\nu}$, is the
sum of the reconstructed muon and kaon energies, $E_{\mu} +
E_{K}$. The \kp and \mum four-vectors are used to calculate the squared
four-momentum transferred to the nucleus:

\begin{equation}
\tcoh = -Q^{2} -2(E_{K}^2 - E_{\nu}p_{K}\cos\theta_{K} + p_{\mu}p_{K}\cos\theta_{\mu K} + m_{K}^{2}),
\end{equation}

\noindent
where $Q^{2}$ is the magnitude of the four-momentum transfer
to the hadronic system, $E_{\mu}$ ($E_{K}$) is the muon (kaon) energy,
$p_{\mu}$ ($p_{K}$) is the magnitude of the muon (kaon)
three-momentum, $\theta_{K}$ is the kaon angle with respect to the
neutrino beam, $\theta_{\mu K}$ is the angle between the outgoing kaon
and muon, and $m_{K}$ is the kaon mass.

The \tcoh\ distributions for data and simulation are shown in
Fig.~\ref{fig:t}. The signal simulation is normalized to the best-fit 
extracted from data. For simulated coherent events, 77\% have reconstructed $\tcoh < 0.1$~GeV$^{2}$ and
94\% have reconstructed $\tcoh < 0.2$~GeV$^{2}$. In simulated signal
events where the kaon energy is underestimated, \tcoh\ is
overestimated. Typically this is due to a \kp\ inelastic interaction in
the detector. The shape in \tcoh\ for background events is due to
available phase space; according to the simulation 83\% of background
events have $\tcoh > 0.2$~GeV$^{2}$. 

The background prediction is constrained by a sideband of events with
reconstructed \tcoh\ between 0.2 and 1.8 GeV$^{2}$. In the simulation,
99.5\% of the events in this region are backgrounds. Prior to the background 
tuning, the simulation overpredicts the background rate, with 
$23.45$ simulated events compared to $13$ observed in data. 
The simulated background is scaled by a factor of 0.55 at all \tcoh. The kaon energy distributions in
simulated background events in the sideband and signal regions are consistent.

\begingroup
\squeezetable
\begin{table}[t]
\begin{tabular}{l|ccc}
\hline\hline
                            & Low $E_{vtx} \tcoh < 1.8$ & $\tcoh < 0.2$   & No \piz (scan) \\
\hline\hline
Best-fit signal             & $4.05$           & $3.80$          & $3.77$ \\
$\Delta S = 0$              & $8.88 \pm 2.58$  & $2.16 \pm 0.65$ & $0.58 \pm 0.20$ \\ 
$\Delta S = 1$ incoherent   & $3.34 \pm 0.97$  & $0.87 \pm 0.26$ & $0.69 \pm 0.21$ \\ 
\kp in detector             & $2.90 \pm 0.84$  & $0.78 \pm 0.23$ & $0.36 \pm 0.12$ \\ 
No \kp                      & $1.89 \pm 0.55$  & $0.30 \pm 0.09$ & $0.14 \pm 0.05$ \\ 
\hline
Total sim. bkg.             & $17.01 \pm 4.94$ & $4.11 \pm 1.20$ & $1.77 \pm 0.53$ \\
Data                        & $21$             & $8$             & $6$ \\
\hline\hline
\end{tabular}
\caption{Counts for events with one \mum and one \kp candidate after each step in the coherent selection. The $20 < E_{vtx} < 60$~MeV cut is applied for all categories. These numbers include the scale factor of $0.55 \pm 0.15$ derived from the high-\tcoh~ sideband. The signal is scaled to the best-fit of the unbinned likelihood fit described in the text. Backgrounds due to ``\kp in detector'' arise when \pip or \kz interact and produce \kp.}
\label{tab:coh_cuts}
\end{table}
\endgroup

\begin{figure}[bt]	
\centering
 \subfloat[]{\includegraphics[width=0.75\columnwidth,bb=10 10 610 310]{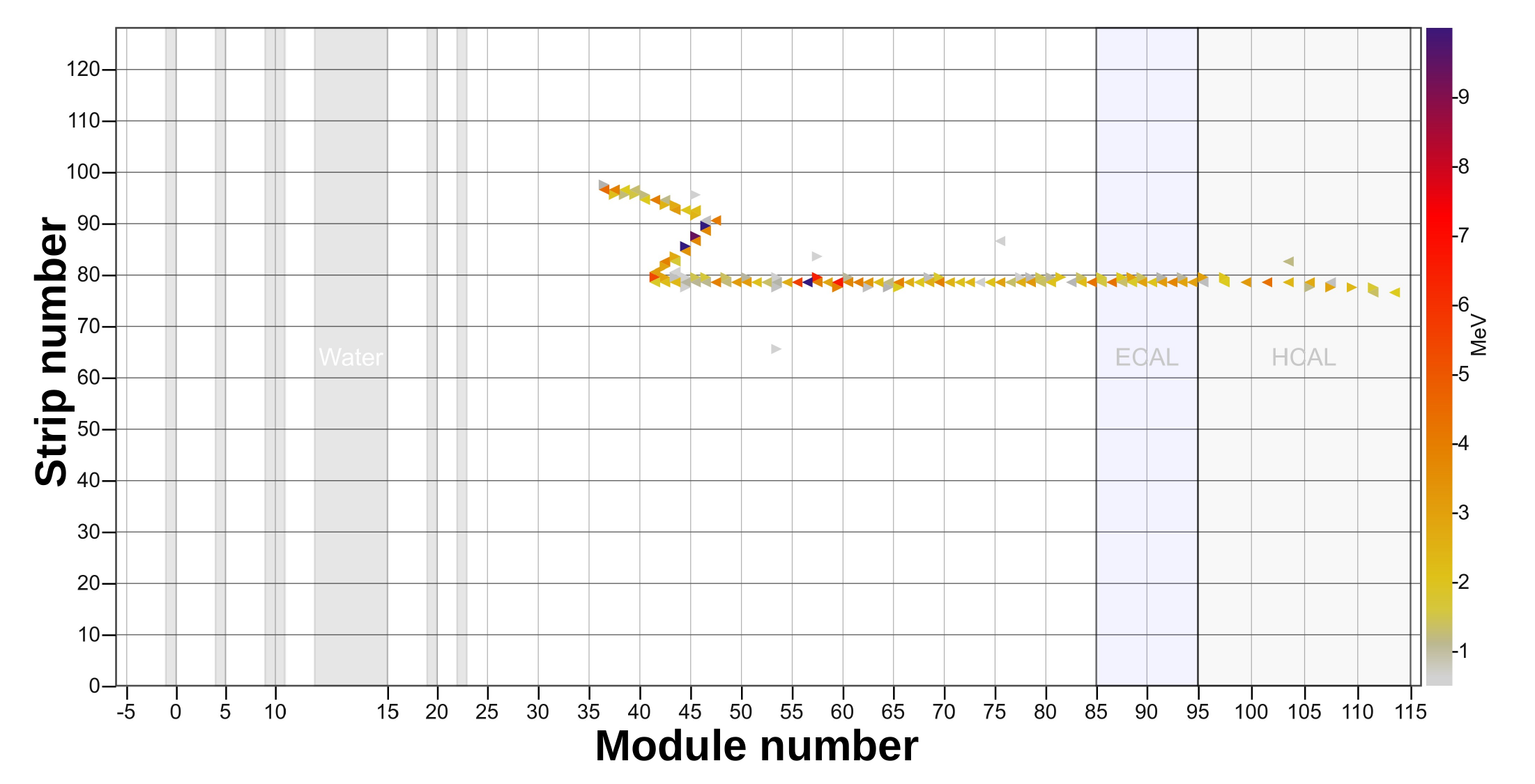} \label{fig:scan-nopi0} } \\
 \subfloat[]{\includegraphics[width=0.75\columnwidth,bb=10 10 610 310]{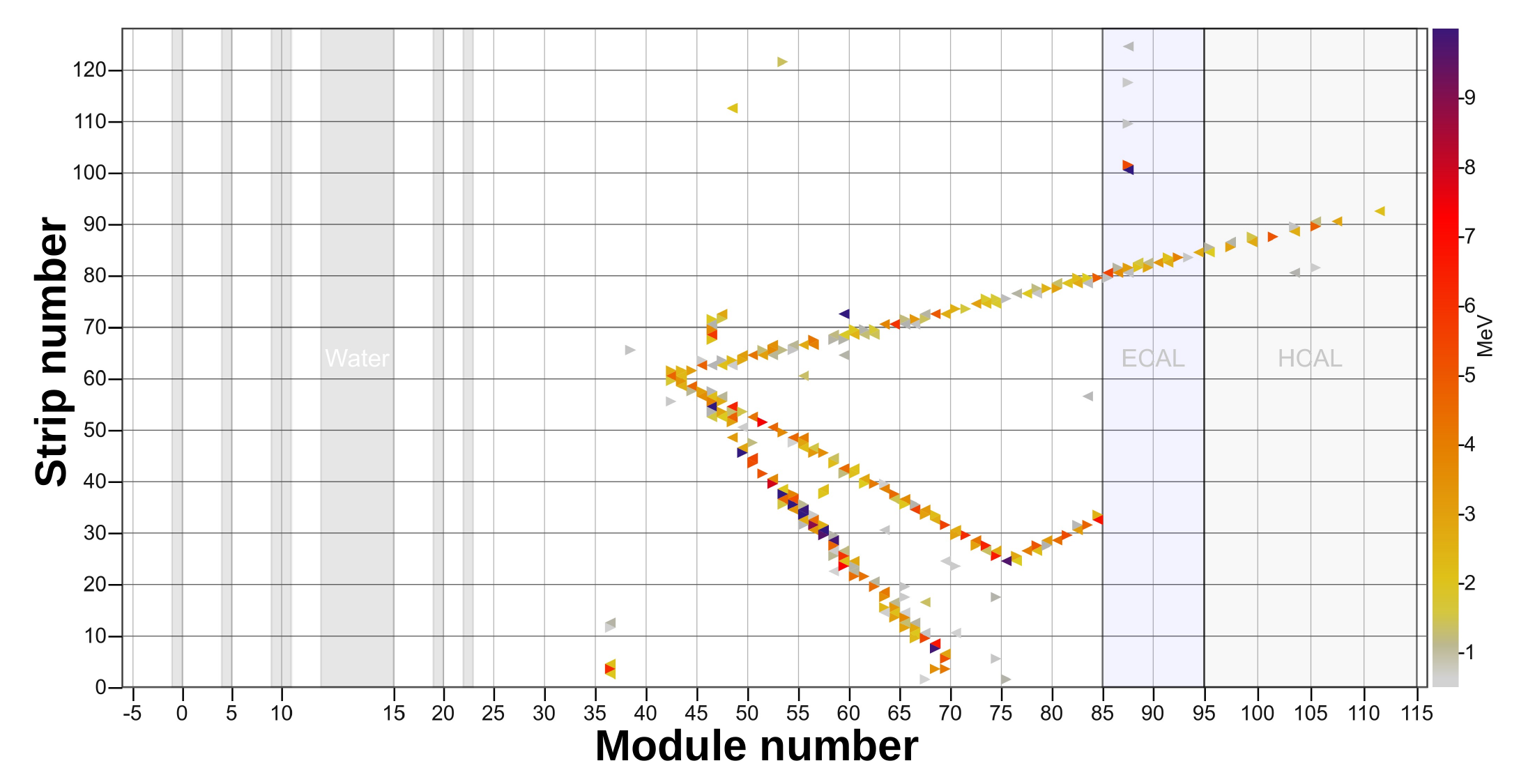} \label{fig:scan-pi0} }
\caption{(a) An event candidate in the data that is not rejected in the scan as having a $\pi^0$ candidate, compared to (b) an event that is rejected. The rejected event clearly has a photon candidate, the widest angle track in the displayed view that is detatched from the neutrino interaction point and points back to the $\mu^-K^+$ vertex and not to a point of an inelastic interaction along the $K^+$ track.  In this event display of hits in the vertically oriented scintillator strips of the detector, the neutrino beam enters from the left, and the tracks that exit out the right of the detector are $\mu^-$.}
\label{fig:pi0-scan}
\end{figure}

Events that satisfy $20 < E_{vtx} < 60$ MeV typically have exactly two charged 
particles emerging from the neutrino interaction point, one \mum and one \kp. 
Background events that pass this cut may also have neutral particles 
that are not observed inside the vertex region. 
The largest single background is
$\numu n \rightarrow \mum \kp \lam$ followed by $\lam \rightarrow
n \piz$. Low vertex energy events are visually scanned to remove
events with an electromagnetic shower which clearly points back to the
neutrino interaction point. The shower is due to the pair conversion
of a photon from the decay $\piz \rightarrow \gamma \gamma$. Examples
of low vertex energy events with and without showers are shown in
Fig.~\ref{fig:pi0-scan}. Events are rejected only if the direction of
the shower can be determined in order to avoid removing events where
the \kp interacts inside the detector and produces a \piz or photon.

Scanning samples contained a mixture of simulated signal and background, and data
events such that the scanner had no knowledge of whether any given
event was from data or simulation. After scanning, 99\% of signal
events are accepted. The small inefficiency is due to \piz produced by \kp
interactions very near the neutrino interaction point. Only 27\% of
$\Delta S = 0$ background events are accepted, along with 80\% of $\Delta S
= 1$ and 46\% of other backgrounds. At low vertex energy, $\Delta S =
1$ events are dominated by $\numu n \rightarrow \mum \kp n$. These
events are rejected when the products of a neutron interaction in the
detector point back to the neutrino interaction point. More often,
the neutron interaction produces a low-energy proton for which a
direction cannot be determined, and the event is accepted. The
scan acceptance is
consistent with no \tcoh\ dependence. Two independent scanners agreed 
on signal efficiency, but differed on $\Delta S = 0$ background efficiencies.
However, the differences applied to both data and simulation, and in the end, 
the two scan results give the same sensitivity to a coherent \kp signal.

A summary of selection cuts is given in Table~\ref{tab:coh_cuts}. The 28\% statistical uncertainty 
on the data in the sideband region and the 8.5\% uncertainty on the integrated flux prediction~\cite{leothesis} are 
fully correlated between the event categories. An uncertainty on the shape of the \tcoh~ distribution is evaluated 
separately for each category by varying parameters in \genie by their uncertainties~\cite{Andreopoulos201087} 
and computing the ratio of the prediction in the sideband and signal regions. The statistical uncertainty on the scan acceptance 
probability is 15\% for $\Delta S = 0$ backgrounds, 6\% for $\Delta S = 1$ incoherent, and 17\% for the other categories.

Diffractive production of \kp from the free protons, $\numu
p \rightarrow \mum \kp p$, would also produce \kp at low \tcoh.
These events have a recoiling proton with kinetic energy $T_p > m_K^4/(8E_K^2m_p)$,
which will most likely leave sufficient energy to fail the vertex energy cut.  
Thus the events are more likely coherent scattering from carbon.

\begin{figure}[tb]
\centering
\mbox{\includegraphics[width=.95\columnwidth,bb=20 20 575 394]{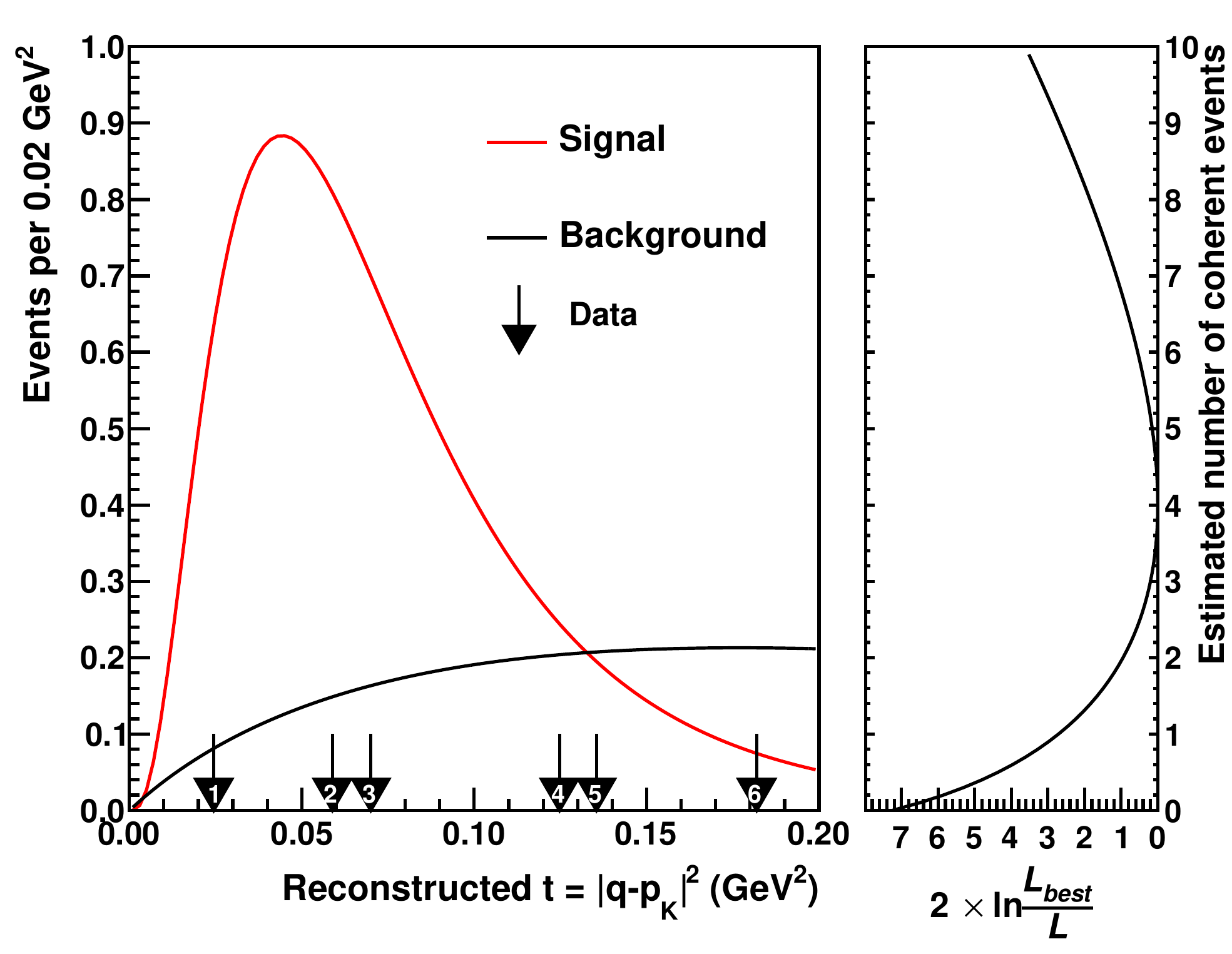}}
\caption{(Left) \tcoh~for selected events after all selections, $\tcoh <0.2$~GeV$^2$ compared to expected distributions for the signal and background. The normalization of the background is fixed by tuning to the high-\tcoh~sideband, while the normalization of the signal is the output of the fit. The data follow the sum of the signal and background. (Right) Twice the log of the ratio of the likelihood at best fit, $L_{best}$, to the likelihood, $L$, as a function of the estimated number of coherent events. This quantity is determined by summing over the six data events and does not include systematic uncertainties. }
\label{fig:t}
\end{figure}
\begin{figure}[tb]
\centering
\mbox{\includegraphics[width=0.85\columnwidth,bb=20 20 575 438]{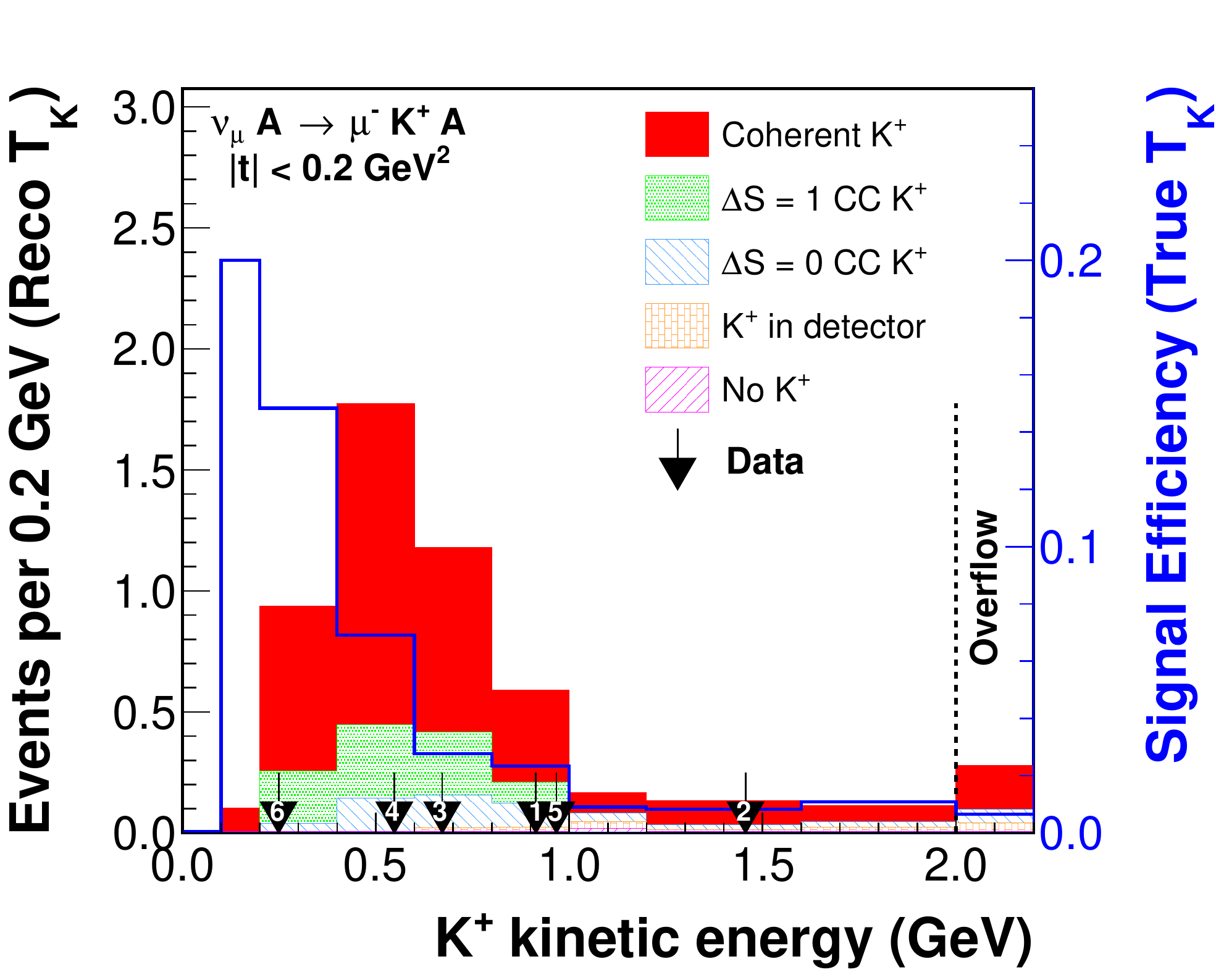}}
\caption{The reconstructed kaon kinetic energy distribution for all accepted events in the $\tcoh <0.2$~GeV$^2$ signal region. Event numbers correspond to the same events in Fig.~\ref{fig:t}. The vertical scale on the right side corresponds to the blue curve, which represents the reconstruction efficiency as a function of true kaon kinetic energy.}
\label{fig:tk}
\end{figure}

We perform an unbinned maximum likelihood fit~\cite{1674-1137-38-9-090001} to the
signal candidates with $\tcoh<0.2$~GeV$^2$. The background normalization 
is fixed within uncertainties by the high-\tcoh~ sideband constraint. 
The only free parameter in the fit is the expected number of signal events for $\tcoh<0.2$~GeV$^2$.
Figure~\ref{fig:t} shows the distribution of $\tcoh$ for these candidate events.
The integrated number of signal events with $\tcoh<0.2$~GeV$^2$ from this fit is
$3.77^{+2.64}_{-1.93}$ events, where the uncertainty is the change to the number of signal events 
required to increase the quantity of twice the log-likelihood by one unit. 
The log-likelihood ratio is shown as a function of the number of coherent events in Fig.~\ref{fig:t}.
We compare the ratio of likelihoods of the null hypothesis of zero signal events to the best fit of $3.77$, 
and find a $p$-value of $0.28$\% including systematic uncertainties, equivalent to a $3.0$ standard deviation exclusion of the null hypothesis
of no coherent kaon production.
The integrated number of predicted background events with $\tcoh < 0.2$~GeV$^2$ is $1.77 \pm 0.53$. Even 
without considering the shape, the observation of six events in data disfavors the null hypothesis.

The number of observed events predicted on the $(1.52\pm0.03)\times 10^{29}$ carbon nuclei in the 
fiducial volume of the detector can be compared with a model
prediction of the cross section using information about the neutrino flux and the acceptance for coherent $K^+$ events, 
where the latter is almost completely determined by the energy of the final state kaon. 
The \kp kinetic energy distribution of selected events with reconstructed $\tcoh < 0.2$ GeV$^2$ is shown in Fig.~\ref{fig:tk}.
The distribution of the six data events is harder than what would be expected for background events. 
This information is not used in the extraction of the $p$-value of $0.28$\% for the background-only hypothesis. 
A total cross section cannot be derived from these data because the total 
acceptance depends strongly on the true kaon energy distribution, which may not be correct in our signal model.

In conclusion, evidence for coherent neutrino production of $K^+$ on carbon nuclei 
has been observed for the first time at $3.0\sigma$ significance by
selecting events with a $\mu^- K^+$ final state,
low momentum transfer to the nucleus, and no evidence of nuclear
breakup. This evidence is
consistent with the Cabibbo-suppressed analog of coherent pion
production, arising from an off-shell $W$ boson converting in the vacuum to a pseudoscalar meson,
and very inconsistent with any other mechanism that does
not have a Cabibbo-suppressed analog.

\begin{acknowledgments}

This work was supported by the Fermi National Accelerator Laboratory
under US Department of Energy contract No. DE-AC02-07CH11359 which
included the \minerva\ construction project.  Construction support was
also granted by the United States National Science Foundation under
Award PHY-0619727 and by the University of Rochester. Support for
participating scientists was provided by NSF and DOE (USA), by CAPES
and CNPq (Brazil), by CoNaCyT (Mexico), by CONICYT (Chile), by
CONCYTEC, DGI-PUCP and IDI/IGI-UNI (Peru), and by Latin American
Center for Physics (CLAF).  One of us (Z.W.) gratefully acknowledges
support from a University of Rochester REACH fellowship.  We thank the
MINOS Collaboration for use of its near detector data. We acknowledge
the dedicated work of the Fermilab staff responsible for the operation
and maintenance of the NuMI beamline, MINERvA and MINOS detectors and
the physical and software environments that support scientific
computing at Fermilab.

\end{acknowledgments}

\bibliographystyle{apsrev4-1}
\bibliography{Coherent}

\clearpage

\newcommand{\qsq}{\ensuremath{Q^2_{QE}}\xspace}
\renewcommand{\textfraction}{0.05}
\renewcommand{\topfraction}{0.95}
\renewcommand{\bottomfraction}{0.95}
\renewcommand{\floatpagefraction}{0.95}
\renewcommand{\dblfloatpagefraction}{0.95}
\renewcommand{\dbltopfraction}{0.95}
\setcounter{totalnumber}{5}
\setcounter{bottomnumber}{3}
\setcounter{topnumber}{3}
\setcounter{dbltopnumber}{3}

\appendix{Appendix: Supplementary Material}\hfill\vspace*{4ex}\

\begingroup
\squeezetable
\begin{table*}[tb]
\centering
\begin{tabular}{l|ccccccccccccc}
\hline \hline
$E_\nu$ (GeV) & 
$0 - 1$ &
$1 - 2$ &
$2 - 3$ &
$3 - 4$ &
$4 - 5$ &
$5 - 6$ &
$6 - 7$ &
$7 - 8$ \\
Flux ($\nu_{\mu}$/cm$^2$/POT ($\times 10^{-9}$)) &
$1.0331  $ &
$4.3611  $ &
$7.4333  $ &
$7.9013  $ &
$3.2984  $ &
$1.2193  $ &
$0.7644  $ &
$0.5671  $ \\
\hline
$E_\nu$ (GeV) & 
$8 - 9$ &
$9 - 10$ &
$10 - 15$ &
$15 - 20$ &
$20 - 25$ &
$25 - 30$ &
$30 - 35$ &
$35 - 40$ \\
Flux ($\nu_{\mu}$/cm$^2$/POT ($\times 10^{-9}$)) &
$0.4398  $ &
$0.3834  $ &
$1.1135  $ &
$0.4664  $ &
$0.1959  $ &
$0.0918  $ &
$0.0668  $ &
$0.0553  $ \\
\hline
$E_\nu$ (GeV) & 
$40 - 45$ &
$45 - 50$ &
$50 - 60$ &
$60 - 70$ &
$70 - 80$ &
$80 - 90$ &
$90 - 100$ &
$100 - 120$ \\
Flux ($\nu_{\mu}$/cm$^2$/POT ($\times 10^{-9}$)) &
$0.0474  $ &
$0.0325  $ &
$0.0267  $ &
$0.0057  $ &
$0.0023  $ &
$0.0006  $ &
$0.0001  $ &
$0.0000  $ \\
\hline
\hline
\end{tabular}
\caption{The predicted $\nu_\mu$ flux per POT for the data included in this analysis.}
\label{tab:nu_flux}
\end{table*}
\endgroup

\begingroup
\squeezetable
\begin{table}[h]
\tabcolsep=0.11cm
\begin{tabular}{l|c|c|c|c|c|c}
& \textcircled{1} & \textcircled{2} & \textcircled{3} & \textcircled{4} & \textcircled{5} & \textcircled{6} \\ 
Reconstructed \tcoh~(GeV$^2$)   & 0.02 & 0.06 & 0.07 & 0.12 & 0.14 & 0.18 \\
Reconstructed $Q^{2}$ (GeV$^2$) & 0.15 & 0.44 & 0.29 & 0.40 & 0.83 & 0.01 \\
Reconstructed $T_{K}$ (GeV)     & 0.91 & 1.46 & 0.67 & 0.97 & 0.55 & 0.25 \\
Reconstructed $E_{\nu}$ (GeV)   & 3.71 & 4.57 & 3.09 & 4.22 & 5.57 & 3.62
\end{tabular}
\caption{The reconstructed squared four-momentum transferred to the nucleus, \tcoh, and to the \kp-nuclus system, $Q^2$, along with the \kp kinetic energy, $T_K$, and neutrino energy, $E_{\nu}$, for the six coherent $K^+$ candidates with $\tcoh <0.2$~GeV$^2$.}
\end{table}
\endgroup

\begingroup
\squeezetable
\begin{table}[h]
\tabcolsep=0.11cm
\begin{tabular}{l|c|c|c|c|c|c|c}
$T_K$ range (GeV) & \textless~0.1 & 0.1-0.2 & 0.2-0.4 & 0.4-0.6 & 0.6-0.8 & 0.8-1.0 & \textgreater~1.0 \\
Acceptance        & 0.00 & 0.20 & 0.15 & 0.07 & 0.03 & 0.02 & 0.01 
\end{tabular}
\caption{Acceptance for signal events as a function of $T_K$.  To a good approximation, the acceptance is independent of the muon or kaon angle produced in our detector.}
\end{table}
\endgroup

\end{document}